\documentclass[letterpaper]{JHEP3}
\usepackage{amssymb,amsfonts}
\usepackage{cite}
\usepackage{epsfig}
\newcommand{\trace}{\mbox{Tr}}

\newcommand{\R}{{\bf R}}

\newcommand{\CD}{{\cal D}}

\newcommand{\CG}{{\cal G}}

\newcommand{\CM}{{\cal M}}
\newcommand{\CN}{{\cal N}}
\newcommand{\CO}{{\cal O}}

\newcommand{\bk}{{\bf k}}
\newcommand{\bx}{{\bf x}}

\newcommand{\p}{\partial}
\renewcommand{\bar}[1]{\overline{#1}}
\renewcommand{\tilde}[1]{\widetilde{#1}}
\newcommand{\be}{\begin{equation}}
\newcommand{\ee}{\end{equation}}
\newcommand{\bea}{\begin{eqnarray}}
\newcommand{\eea}{\end{eqnarray}}

\usepackage{graphicx}
\usepackage{latexsym}

\newcommand{\ie}{{\it i.e.}}
\newcommand{\eg}{{\it e.g.}}
\newcommand{\gym}{g_{\rm YM}}
\newcommand{\geff}{g_{\rm eff}}
\title{Quantum Criticality and Yang-Mills Gauge Theory}
\author{Petr Ho\v{r}ava\\
Berkeley Center for Theoretical Physics and Department of Physics\\
University of California, Berkeley, CA, 94720-7300\\
and\\
Theoretical Physics Group, Lawrence Berkeley National Laboratory\\
Berkeley, CA 94720-8162, USA}
\abstract{We present a family of nonrelativistic Yang-Mills gauge theories 
in $D+1$ dimensions whose free-field limit exhibits quantum critical behavior 
with gapless excitations and dynamical critical exponent $z=2$.  The ground 
state wavefunction is intimately related to the partition function of 
relativistic Yang-Mills in $D$ dimensions.  
The gauge couplings exhibit logarithmic scaling and asymptotic freedom in the 
upper critical spacetime dimension, equal to $4+1$.  The theories can be 
deformed in the infrared by a relevant operator that restores Poincar\'e 
invariance as an accidental symmetry.  In the large-$N$ limit, our 
nonrelativistic gauge theories can be expected to have weakly curved gravity 
duals.}  
\begin{document}


We present a class of nonrelativistic Yang-Mills gauge theories which exhibit 
anisotropic scaling between space and time.  Our motivation originates from 
several different areas of physics, which have been experiencing a stimulating 
confluence of theoretical ideas recently: condensed matter theory, in 
particular quantum critical phenomena, string theory, and gauge-gravity 
duality.  

In our study, we consider the case of $D+1$ spacetime dimensions, continuing 
the viewpoint advocated in \cite{kfermi} that the interface of condensed 
matter and string theory is best studied from the vantage point of arbitrary 
number of dimensions, even though practical applications to condensed matter 
are likely to be expected only for $D\leq 3$.  

The theories presented here are candidates for the description of new 
universality classes of quantum critical phenomena in various dimensions.  
They combine the idea of non-Abelian gauge symmetry, mostly popular in 
relativistic high-energy physics, with the concept of scaling with 
non-relativistic values of the dynamical critical exponent, $z\neq 1$.  
\footnote{A different proposal for a quantum critical gauge theory, in $2+1$ 
dimensions and with $\CG=SU(2)$, was made in \cite{freedman}.}
In combination, this anisotropic scaling together with Yang-Mills symmetry 
opens up a new perspective on gauge theories, changing some of the basic 
features of relativistic Yang-Mills such as the critical dimension in which 
the theory exhibits logarithmic scaling.

Since our theories can be constructed for any compact gauge group, the choice 
of the $SU(N)$, $SO(N)$ or $Sp(N)$ series yields a class of theories with 
anisotropic scaling and a natural large-$N$ expansion parameter. These 
theories can then be expected to have weakly-curved gravitational duals,  
perhaps leading to new realizations of the AdS/nonrelativistic CFT 
correspondence which has attracted considerable attention recently 
\cite{son,mcg,talk,kachru}.  Finally, it turns out that our theories are 
intimately related to relativistic theories in one fewer dimension, and 
therefore can shed some new light on the dynamics of the relativistic models.  

\section{Theories of the Lifshitz Type}

We work on a spacetime of the form $\R\times\R^D$, with coordinates $t$ and 
$\bx\equiv (x^i)$, $i=1,\ldots D$, equipped with the flat spatial metric 
$\delta_{ij}$ (and the metric $g_{tt}=1$ on the time dimension).  The theories 
proposed here are of the Lifshitz type, and exhibit fixed points with 
anisotropic scaling characterized by dynamical critical exponent $z$ 
(see, \eg, \cite{sachdev}),
\be
\bx\rightarrow b\bx,\qquad t\rightarrow b^z t.
\ee
We will measure dimensions of operators in the units of spatial momenta, 
defining 
\be
[x^i]=-1,\qquad [t]=-z.
\ee
The prototype of a quantum field theory with nontrivial dynamical exponent 
$z$ is the theory of a single Lifshitz scalar $\phi(\bx,t)$.  In its simplest 
incarnation, this theory is described by the following action, 
\be
\label{lifaction}
S=\frac{1}{2}\int dt\,d^D\bx\left\{(\dot\phi)^2-\frac{1}{4\kappa^2}
(\Delta\phi)^2\right\},
\ee
where $\Delta\equiv\p_i\p_i$ is the spatial Laplacian.  Throughout most of the 
paper we adhere to the nonrelativistic notation, and denote the time 
derivative by ``$\dot{\ \ }$''.

The Lifshitz scalar is a free-field fixed point with $z=2$.  The engineering 
dimension of $\phi$ is $[\phi]=D/2-1$, \ie , the same as the dimension of the 
relativistic scalar in $D$ spacetime dimensions, implying an interesting shift 
in the critical dimensions of the $z=2$ system compared to its relativistic 
cousin.

Note that the potential term in the Lifshitz action (\ref{lifaction}) is of 
the form 
\be
\left(\frac{\delta W[\phi]}{\delta\phi}\right)^2,
\ee
where $W$ is the Euclidean action of a massless relativistic scalar in $D$ 
dimensions,
\be
\label{euclif}
W[\phi]=\frac{1}{2\kappa}\int d^D\bx\,(\p_i\phi\p_i\phi).
\ee
When Wick rotated to imaginary time $\tau=it$, the action can be written as a 
perfect square, 
\be
\label{lifactsq}
S=\frac{i}{2}\int d\tau\,d^D\bx\left\{\left(\p_\tau\phi+\frac{1}{2\kappa}
\Delta\phi\right)^2\right\},
\ee
because the cross-term in (\ref{lifactsq}) is a total derivative, 
$\dot\phi\,\Delta \phi/\kappa=-\dot W$, and can be dropped.

The coupling $\kappa\in[0,\infty)$ parametrizes a line of fixed points.  
If we wish, we can absorb $\kappa$ into the rescaling of the time coordinate 
and a rescaling of $\phi$.  

In the original condensed-matter applications 
\cite{hornreich,ardonne,lubensky}, the anisotropy is between 
different spatial dimensions, and the Lifshitz scalar is designed to describe 
the tricritical point at the juncture of the phases with a zero, homogeneous 
and spatially modulated condensate.  

\section{Yang-Mills Theory and Quantum Criticality}

Our nonrelativistic gauge theory in $D+1$ dimensions will be similarly 
associated with with relativistic Yang-Mills in $D$ dimensions.  

Our gauge field is a one-form on spacetime, with spatial components 
$A_i=A^a_i(x^j,t)T_a$ and a time component $A_0=A^a_0(x^i,t)T_a$.  The Lie 
algebra generators $T_a$ of the gauge group $\CG$ (which we take to be compact 
and simple or a $U(1)$) satisfy commutation relations 
$[T_a,T_b]=if_{ab}{}^cT_c$.  We normalize the trace on the Lie 
algebra of $\CG$ by $\trace(T_aT_b)=\frac{1}{2}\delta_{ab}$.  

Our theory will be invariant under gauge symmetries 
\be
\delta_\epsilon A_i=\left(\p_i\epsilon^a+f_{bc}{}^aA_i^b\epsilon^c\right)T_a
\equiv D_i\epsilon,\qquad \delta_\epsilon A_0=\dot\epsilon -i[A_0,\epsilon].  
\ee
Gauge-invariant Lagrangians will be constructed from the field strengths
\bea
E_i&=&\left(\dot A^a_i-\p_i A^a_0 +f_{bc}{}^aA^b_iA^c_0\right)T_a=
\dot A_i-\p_i A_0-i[A_i,A_0],\cr
F_{ij}&=&\left(\p_iA^a_j-\p_iA^a_i+f_{bc}{}^aA^b_iA^c_j\right)T_a
=\p_iA_j-\p_iA_i-i[A_i,A_j].
\eea
We will now construct a theory which has $z=2$ in the free field limit.  
The engineering dimensions of the gauge field components at the corresponding 
Gaussian fixed point will be 
\be
[A_i]=1,\qquad [A_0]=2.
\ee

The Lagrangian should contain a kinetic term which is quadratic in first 
time derivatives, and gauge invariant.  The unique candidate for this 
kinetic term is $\trace(E_iE_i)$, of dimension $[\trace(E_iE_i)]=6$.  One can 
then follow the strategy of effective field theory, and add all possible 
terms with dimensions ${}\leq 6$ to the Lagrangian.  This would allow 
terms such as $\trace(F_{ij}F_{k\ell}F_{\ell i})$, 
$\trace(D_iF_{jk}D_iF_{jk})$, $\trace(D_iF_{ik}D_jF_{jk})$, (all of dimension 
six), a term $\trace(F_{ij}F_{ij})$ of 
dimension four, etc.  One could indeed define the theory in this fashion, 
study the renormalization-group (RG) behavior in the space of all the 
couplings, and look for 
possible fixed points.  This interesting problem is beyond the scope of the 
present paper.  Instead, we pursue a different strategy, and limit the number 
of independent couplings in a way compatible with renormalization.  The trick 
that we will use is familiar from a variety of areas of physics, such as 
dynamical critical systems \cite{halperin,ma}, stochastic quantization 
\cite{parisi,namiki}, and nonequilibrium statistical mechanics.  

Inspired by the structure of the Lifshitz scalar theory, we take our action 
to be
\be
\label{actn}
S=\frac{1}{2}\int dt\,d^D\bx\left\{\frac{1}{e^2}\trace(E_iE_i)-\frac{1}{g^2}
\trace\left((D_iF_{ik})(D_jF_{jk})\vphantom{\tilde E}\right)\right\}.
\ee
This is a Lagrangian with $z=2$ and no Galilean invariance.  As a result, 
there is no symmetry relating the kinetic term and the potential term, and 
therefore no {\it a priori\/} relation between the renormalization of the two 
couplings $e$ and $g$.  The potential term is again the square of the equation 
of motion that follow from an action:  the relativistic Yang-Mills in $D$ 
Euclidean dimensions.  When a theory in $D+1$ dimensions is so constructed 
from the action of a theory in $D$ dimensions, we will say that it 
{\it satisfies the detailed balance condition}, borrowing the terminology 
common in nonequilibrium dynamics.  

\section{At the Free-Field Fixed Point with $z=2$}

The free-field fixed point will be obtained from (\ref{actn}) by taking 
$e$ and $g$ simultaneously to zero.  Keeping both the kinetic and the 
potential term finite in this limit requires rescaling the gauge field,  
$\tilde A^a_i\equiv A^a_i/\sqrt{eg}$, and keeping $\tilde A_i^a$ finite as we 
take $e$ and $g$ to zero.  This gives
\be
\label{actnff}
S=\frac{1}{2}\int dt\,d^D\bx\left\{\frac{g}{e}\trace(\tilde E_i\tilde E_i)
-\frac{e}{g}\trace\left((\p_i\tilde F_{ik})(\p_j\tilde F_{jk})\right)\right\},
\ee
where $\tilde E_i$ and $\tilde F_{ij}$ are the linearized field strengths of 
$\tilde A_i$.  

We see that there is actually a line of free fixed points, parametrized by the 
dimensionless ratio 
\be
\lambda=\frac{g}{e}.
\ee
As in the Lifshitz scalar theory, if we wish we can absorb $\lambda$ 
into a rescaling of time, $t_{\rm new}=t/\lambda$.  

The special properties of the Lifshitz scalar make it possible to determine 
the exact ground-state wavefunction \cite{ardonne}, 
\be
\Psi[\phi(\bx)]=\exp\left\{-\frac{1}{4\kappa}\int d^D\bx (\p_i\phi\,\p_i\phi)
\right\}.
\ee
This $\Psi$ is equal to $\exp(-W[\phi]/2)$, where $W[\phi]$ is the action 
(\ref{euclif}) of the relativistic scalar in $D$ dimensions.  The norm 
$\int\CD\phi(\bx)\,\Psi^\ast\Psi$ equals the partition function of this 
relativistic theory.  

Similarly, we can relate the ground-state wavefunction of our $z=2$ gauge 
theory to the partition function of relativistic Yang-Mills.  
The momenta and the Hamiltonian are
\be
\tilde P_i^a=\frac{\lambda}{2}\tilde E_i^a,\qquad H=\frac{2}{\lambda}
\int dt\,d^D\bx\,\trace\left\{\tilde P_i\tilde P_i+\frac{1}{4}(\p_i\tilde 
F_{ik})(\p_j\tilde F_{jk})\right\}. 
\ee
The preferred role of time suggests a natural gauge-fixing condition,
\be
\label{gauge}
A_0=0.
\ee
This does not fix all gauge symmetries, leaving the subgroup of 
time-independent gauge transformations unfixed.  We can eliminate 
the residual gauge invariance by setting
\be
\label{secgf}
\p_iA_i=0
\ee
at some fixed time slice with $t=t_0$.  However, since the equations of 
motion obtained from varying $A_0$ yield $\p_i\dot A_i-D_iD_iA_0=0$,
once we adopt the gauge (\ref{gauge}) and select (\ref{secgf}) at $t=t_0$,   
this condition will continue to hold for all $t$.  First we consider the 
linearized theory describing the free-field fixed point.  The Hamiltonian 
operator can be written as
\be
H=\frac{1}{\lambda}\int dt\,d^D\bx\,\trace(Q^\dagger Q+QQ^\dagger)
=\frac{2}{\lambda}\int dt\,d^D\bx\,\trace(Q^\dagger Q)+E_0,
\ee
where we have defined
\be
Q_i^a=-\frac{\delta}{\delta\tilde A_i^a}+\frac{1}{2}\p_k\tilde F^a_{ki}.
\ee
The vacuum energy $E_0$ is a field-independent normal-ordering 
constant.  The energy will be minimized by solutions of $Q\Psi[A_i]=0$.  
Thus, we obtain the ground-state wavefunction 
\be
\label{grst}
\Psi[A_i(\bx)]=\exp\left\{-\frac{1}{4}\int\trace (\tilde F_{ij}\tilde F_{ij})
\right\}=\exp\left\{-\frac{1}{4eg}\int\trace (F_{ij}F_{ij})\right\},
\ee
where we have restored the original normalization of the linearized gauge 
field.  The exponent is one half of the quadratic part of the Euclidean 
Yang-Mills action
\be
W[A_i]=\frac{1}{2\gym^2}\int d^D\bx\,\trace(F_{ij}F_{ij})
\ee
in Lorenz gauge, if we identify the Yang-Mills coupling as
\be
\gym^2=eg.
\ee
A closer inspection shows that (\ref{grst}) is indeed the correct ground-state 
wavefunction of the linearized theory.  In particular, the energy is bounded 
from below, and the spectrum of excitations at the free-field fixed point 
consists of $D-1$ polarizations of gauge bosons with the gapless 
nonrelativistic dispersion relation 
\be
\omega^2=\frac{(\bk^2)^2}{\lambda^2}, 
\ee
Note the presence of both positive and negative frequency modes: The concept 
of particles and antiparticles in our model is similar to that of a 
relativistic theory.  

The form of the ground-state wavefunction (\ref{grst}) should be contrasted 
with that of relativistic Yang-Mills theory in $3+1$ dimensions, which also 
obeys the detailed balance condition, with the Chern-Simons action
\be
W_{\rm CS}=\int\omega_3(A),\qquad \omega_3(A)=\trace\left(A\wedge dA
+\frac{2}{3}A\wedge A\wedge A\right).
\ee
This makes $\Psi\sim\exp(-W_{\rm CS}/2)$ a candidate solution of the 
Schr\"odinger equation.  This solution is unphysical, for many reasons 
(see \cite{witten} for a detailed discussion), while in our case the solution 
(\ref{grst}) of the Schr\"odinger equation is the physical ground-state 
wavefunction.  

The anisotropic scaling with $z=2$ shifts the critical dimension of the 
system:  The engineering dimensions of the couplings at our free fixed point 
are
\be
[e^2]=[g^2]=4-D.
\ee
The system is at its upper critical dimension when $D=4$, \ie , in $4+1$ 
spacetime dimensions.  

\section{Quantization}

The quantum theory is formally defined by the path-integral 
\be
\int \CD A_i\,\exp(iS[A]),
\ee
where $\CD A_i$ is the gauge-invariant measure on the space of gauge orbits.  
As in relativistic quantum field theory, it will be convenient 
to perform the Wick rotation to imaginary time, $\tau=it$.  After the Wick 
rotation, our action can be rewritten in the following form,
\bea
\label{sqact}
S&=&\frac{i}{2}\int d\tau\,d^D\bx\,\trace\left(\frac{1}{e^2}E_kE_k
+\frac{1}{g^2}D_iF_{ik}D_jF_{jk}\right)\cr
&&\qquad{}=\frac{i}{2}\int d\tau\,d^D\bx\,\trace\left\{\left(\frac{1}{e}E_k
+\frac{1}{g}D_iF_{ik}\right)\left(\frac{1}{e}E_k+\frac{1}{g}D_jF_{jk}\right)
\right\},  
\eea
\ie , as a sum of squares, one for each field $A_i^a$.  The cross-terms are a 
combination of a total derivative, $\dot A_iD_kF_{ik}/(2eg)=\dot W[A_j]$, and 
a gauge transformation, $(D_iA_0)(D_kF_{ik})/(2eg)=\delta_{A_0}W[A_j]=0$, of 
the $D$-dimensional Euclidean Yang-Mills action $W[A_i]$.  The imaginary-time 
action (\ref{sqact}) is formally equivalent to the action that appears in 
stochastic quantization \cite{parisi} of the relativistic Yang-Mills theory 
in $D$ dimensions, if we identify $\tau$ as the ``fictitious time,'' and $A_0$ 
as the field introduced in the process of ``stochastic gauge fixing''  
\cite{zinnzwan}.

Using an auxiliary field $B^a_j$, we can rewrite
\be
\int\CD A_i\,\exp(iS)=\int\CD A_i\,\CD B_i\,\exp\left\{-\int d\tau\,d^D\bx
\left(B_i\left(\frac{1}{e}E_i+\frac{1}{g}D_kF_{ik}\right)-\frac{1}{2}B_iB_i
\right)\right\}.
\ee
The key to the renormalizability of this setup \cite{zinnzwan} is to show that 
any field-dependent counterterm generated is at least linear in $B^a_i$.%
\footnote{An important part of the argument is to show that the Jacobian of 
the change of variables from $A_i$ to $\dot A_i+D_kF_{ik}/\lambda$ is 
independent of $A_k$, see \cite{zinnzwan} for details.  One can also represent 
the Jacobian by integrating in a pair of fermions $\eta_i^a$, $\bar\eta_i^a$.  
Together with $A_i^a$ and $B_i^a$, these fields can be interpreted as members 
of a supermultiplet of Parisi-Sourlas supersymmetry.  We will not use such a 
supersymmetric framework in this paper.}
As a result, the theory inherits the renormalization properties from the 
associated theory in $D$ dimensions, with the only new renormalization to be 
performed being that of the relative normalization of time and space.  

In the critical dimension $4+1$, this ``quantum inheritance principle'' 
suggests that our nonrelativistic gauge theory could be asymptotically free.  
We define the RG beta functions  $\beta_e=\mu(d/d\mu)e(\mu)$, $\beta_g=
\mu(d/d\mu)g(\mu)$, with $e(\mu)$ and $g(\mu)$ the renormalized couplings 
and $\mu$ the RG scale.  Borrowing results from stochastic quantization, we 
get at one loop in $\gym$ \cite{bern,okano,namiki}
\be
\label{betas}
\beta_e=-\frac{3}{2}C_2\,e^2g+\ldots,\qquad
\beta_g=-\frac{35}{6}C_2\,eg^2+\ldots,
\ee
where $C_2\equiv c_2(\CG)/(4\pi)^2$, with $c_2(\CG)$ the quadratic Casimir of 
the adjoint of $\CG$ (for example, $c_2(SU(N))=N$).  This RG flow pattern 
implied by (\ref{betas}) 
can be disentangled by switching from $(e,g)$ to $(\gym,\lambda)$, which 
yields
\be
\mu\frac{d\gym}{d\mu}=-\frac{11}{3}C_2\,\gym^3+
\CO\left(\gym^5\right),\qquad
\mu\frac{d\lambda}{d\mu}=-\frac{13}{3}C_2\,\gym^2\lambda+\CO\left(
\gym^4\lambda\right).
\ee
This is solved by
\be
\frac{1}{\gym^2(\mu)}=\frac{1}{\gym^2(\mu_0)}+\frac{22}{3}C_2\,
\log\left(\frac{\mu}{\mu_0}\right),\qquad\lambda(\mu)=\lambda(\mu_0)\left(
\frac{\gym(\mu)}{\gym(\mu_0)}\right)^{13/11}.
\ee
The theory is indeed asymptotically free.  Moreover, the one-loop beta 
function for $\gym$ coincides with that of relativistic Yang-Mills in four 
dimensions.  The theory exhibits dimensional transmutation, with a dynamically 
generated scale $\Lambda$.  The line of Gaussian fixed points parametrized by 
$\lambda$ is lifted: Nonzero $\gym$ induces a flow of $\lambda$ towards 
smaller values.  

In lower dimensions $D<4$, the theory superrenormalizable by power couting,  
and we expect it to become strongly coupled in the infrared (IR).  It would be 
interesting to investigate this regime further, with possible applications to 
quantum critical systems in mind.  Gauge-gravity duality might be the right 
technique how to understand such strongly coupled theories, especially at 
large $N$.  

\section{Relevant Deformations: Softly Broken Detailed Balance}

The asymptotic freedom of our theory in $4+1$ dimensions suggests the 
possibility of using it as a UV completion of otherwise nonrenormalizable 
effective theories, and in particular, of relativistic Yang-Mills theory.  

In theories that satisfy the detailed balance condidition, we 
could add relevant deformations to $W$, preserving the detailed balance 
condition.  For example, adding $m_0^2\phi^2$ to $W[\phi]$ of the Lifshitz 
scalar theory is such a deformation; the theory flows under this deformation 
to a $z=1$ theory with accidental relativistic invariance in the nfrared.  
Our Yang-Mills theory has no relevant deformations of this type.  

Alternatively, we can add relevant deformations directly to $S$.  This will 
represent a soft violation of the detailed balance condition.   There is one 
relevant term that can so be added to $S$, 
\be
\Delta S=-m^{D-2}\int dt\,d^D\bx\,\trace(F_{ij}F_{ij}).
\ee
If $m^{D-2}>0$, this will lead naturally to a $z=1$ relativistic theory 
at long distances.  No other, lower-dimensional nontrivial gauge-invariant 
scalar operators exist.   

In order to make the relativistic symmetry in the infrared manifest, it is 
natural to rescale the time dimension,
\be
x^0=ct.  
\ee
In terms of the UV variables, the speed of light is given by
\be
c=2m^{D/2-1}e.
\ee
Until we couple matter to the Yang-Mills sector, the relativistic symmetry is 
protected by the gauge invariance if the gauge group is simple.  In terms of 
the relativistic notation for the fields, $A_\mu=(A_0/c,A_i)$, our deformed 
classical action is
\be
\label{actrel}
S+\Delta S=-\frac{1}{2\geff^2}\int d^{D+1}x\,\left\{
\trace(F_{\mu\nu}F^{\mu\nu})+\frac{m^{2-D}}{2g^2}\trace\left((D_kF_{ki})
(D_jF_{ji})\vphantom{\tilde E}\right)\right\},
\ee
The dimensionful effective Yang-Mills coupling of the infrared theory is given 
by
\be
\geff^2=m^{1-D/2}e.
\ee
From the perspective of the infrared free-field fixed point, the $(DF)^2$ 
term in (\ref{actrel}) represents an irrelevant deformation, which modifies 
the relativistic massless dispersion relation 
$k_\mu k^\mu\equiv -k_0^2+\bk^2=0$ to 
\be
k_\mu k^\mu=-\frac{m^{2-D}}{2g^2}\,(k_ik_i)^2.
\ee
This correction only becomes important at high energies, confirming that 
microscopically there is no speed limit in this theory. In the relativistic 
spacetime coordinates, the dispersion relation asymptotes at large $\bk$ to 
$|k_0|=\bk^2/(2\CM)$, with $\CM=m^{D/2-1}g$.

In the free-field limit $\gym=0$, this flow from the $z=2$ UV fixed point to 
the relativistic $z=1$ theory in the infrared is exact.  Turning on $\gym$ 
in $4+1$ dimensions leads to dimensional transmutation, implying a competition 
of scales:  If $m\gg\Lambda$, the theory starts flowing towards the $z=1$ 
IR fixed point before reaching strong coupling, while for $m\ll\Lambda$ it is 
driven to strong coupling first.  

Another interesting possibility is to start with $m^{D-2}<0$, which would be 
analogous to the spatially modulated phases of the Lifshitz scalar theory.  

\section{Conclusions}

In this paper, we focused on the basic features of the simplest, bosonic 
version of Yang-Mills gauge theory with anisotropic scaling in $D+1$ 
dimensions. Clearly, many interesting questions remain open for further 
study.  

One immediate question is whether this framework can be supersymmetrized, 
at least in the $D$ dimensional sense:  One can consider replacing the 
relativistic $D$ dimensional theory in our construction with one of its 
supersymmetric extensions.  If the quantum inheritance principle continues to 
hold in such cases, our construction might lead to new nontrivial 
nonrelativistic RG fixed points in $D+1$ dimensions.  One particularly 
tempting question to ask is how the $\CN=4$ super Yang-Mills CFT fits 
into this framework.  

It should also be interesting to see whether the class of nonrelativistic 
gauge theories presented in this paper can be engineered from string theory, 
in particular from D-brane configurations.  Such a construction would allow 
a more systematic study of these theories and their anticipated 
gravitational duals.  

\vfill\break
\acknowledgments
This work has been supported by NSF Grant PHY-0555662, DOE Grant 
DE-AC03-76SF00098, and the Berkeley Center for Theoretical Physics.  

\bibliographystyle{JHEP}
\bibliography{cym}
\end{document}